\documentclass{aa}

\usepackage{graphicx}
\usepackage{amsmath}
\usepackage{amssymb}
\usepackage{natbib}
\usepackage{tabularx}
\usepackage{longtable}
\usepackage{url}
\bibpunct{(}{)}{;}{a}{}{,}

\makeatletter
\let\makeLineNumber\relax
\makeatother

\begin{document}

\title{From stardust to interstellar grain growth in the first galaxies:
a cosmological transition in dust evolution near $z \approx 8.9$}
\author{D. Burgarella\inst{1}\thanks{Corresponding author: denis.burgarella@lam.fr}
\and T. J. L. C. Bakx\inst{2}
\and R. Schneider\inst{3,4,5,6}
\and H. S. B. Algera\inst{7}
\and C. Aurin\inst{1}
\and T. Dewachter\inst{1}
\and C. Esmerian\inst{2}
\and L. Graziani\inst{3,4,5}
\and K. Knudsen\inst{2}
\and K. Otaki\inst{9}}

\institute{Aix Marseille Universit\'e, CNRS, CNES, LAM, Marseille, France
\and Department of Physics and Astronomy, Chalmers University of Technology, SE-412 96 Gothenburg, Sweden
\and Dipartimento di Fisica, ``Sapienza'' Universit\`a di Roma, Piazzale Aldo Moro 5, 00185 Roma, Italy
\and INAF/Osservatorio Astronomico di Roma, Via Frascati 33, 00040 Monte Porzio Catone, Italy
\and INFN, Sezione di Roma I, P.le Aldo Moro 2, I-00185 Roma, Italy
\and Sapienza School for Advanced Studies, Viale Regina Elena 291, 00161 Roma, Italy
\and Institute of Astronomy and Astrophysics, Academia Sinica, 11F of Astronomy-Mathematics Building, No.1, Sec. 4, Roosevelt Rd, Taipei 106319, Taiwan, R.O.C.
\and Information Technology Center, The University of Tokyo, 6-2-3 Kashiwanoha, Kashiwa, Chiba 277-0882, Japan}

\date{Received -- / Accepted --}

\abstract
{When and how did dust begin to shape galaxies? Motivated by the identification of an apparent redshift break in galaxy dust masses, suggesting substantially lower dust masses at $z \gtrsim 9$, we investigate dust enrichment during the first billion years of cosmic history.}
{We aim to determine whether the observed evolution marks a transition in the dominant dust-production mechanism and to identify the physical conditions under which such a transition is expected to occur.}
{Using JWST, ALMA, and NOEMA observations, we measure ultraviolet dust attenuation and dust masses. We apply a censored change-point analysis and compare the observations with dust evolution modelling.}
{The analysis identifies a preferred transition near redshift $z \simeq 8.9$, corresponding to about 570 Myr after the Big Bang. The evidence for a break is strongest in dust mass and dust-to-stellar mass ratio, mostly estimated from JWST NIRSpec spectrophotometric fitting but also partly from sub-mm data. The ultraviolet attenuation measurements are consistent with a transition at the same epoch but do not independently require one. The models are consistent with the onset of efficient interstellar grain growth above a characteristic metallicity.}
{We interpret the transition near $z \simeq 9$ as the emergence of grain-growth-dominated dust evolution from an earlier regime dominated by supernova-produced grains. Population III enrichment can modify the earliest chemical-enrichment history but leaves the timing of the dust transition nearly unchanged and is not required for its emergence.}


\maketitle

\section{Introduction}
\label{sec:introduction}

Dust plays a central role in shaping galaxies, yet its origin in the early Universe remains uncertain \citep[e.g.][]{ref01,ref02,ref03}. Observations with the Atacama Large Millimeter/submillimeter Array (ALMA) have detected substantial dust masses in galaxies at redshifts up to $z \simeq 8$ \citep{ref04,ref05,ref06,ref07,ref08,ref09,ref10}. At the
same time, recent observations with the James Webb Space Telescope (JWST) have uncovered a population of galaxies at $z \gtrsim 9$ with extremely low dust attenuation (GELDAs), suggesting that some early systems contain very little dust \citep{ref11,ref12,ref13}. Independent ALMA measurements \citep{ref15,ref16}, as well as far-ultraviolet (FUV) slopes $\beta_{FUV}$ estimated from JWST data \citep{ref17}, support this interpretation. Their observational appearance initially led to the phenomenological label ``Blue Monsters'' \citep{ref13,ref14}, without in itself identifying the physical origin of the underlying phenomenon. 

The coexistence of dusty galaxies at $z \simeq 8$ and dust-poor systems at $z > 9$ suggests a population level change in the typical dust content of early galaxies. The central question asked here is whether this change reflects a transition in the dominant dust production mechanism. Dust injected by supernovae (SNe) can seed the interstellar medium (ISM), but the large dust masses observed in galaxies are difficult to explain without subsequent grain growth in the dense ISM \citep{ref01,ref02,ref03,ref12}. Because grain growth strongly depends on metallicity, dust enrichment may accelerate once galaxies reach a critical metal enrichment threshold ($Z_{crit}$).

Because dust regulates cooling, chemistry, and the escape of UV
radiation, identifying when this transition occurred is essential for understanding galaxy evolution during cosmic dawn.

This paper combines JWST measurements of UV dust attenuation
($A_{\rm FUV}$) with ALMA dust mass ($M_{\rm dust}$) constraints to investigate the origin of dust grains during cosmic dawn. Using dust evolution models, we show that this transition arises naturally when galaxies cross the critical metallicity required for efficient ISM grain growth \citep{ref18, ref41}.

\section{Methodology used in this paper}
\label{sec:methodology}

\subsection{Observational samples}
\label{subsec:samples}

We combine three published samples that provide complementary
constraints on dust attenuation and dust mass during cosmic dawn
\footnote{The catalogs of derived physical properties used in this work are publicly available at: Burgarella et al. (2025), \url{https://amubox.univ-amu.fr/s/YNyWABox6TegndX}; Burgarella et al. (2026), \url{https://amubox.univ-amu.fr/s/kbwJWgmCekn4BnS}; and Bakx et al. (2026), \url{https://academic.oup.com/mnras/article/546/2/staf2284/8407241?login=false}.}

The first sample is drawn from the JWST/CEERS NIRSpec survey and consists of 173 galaxies with secure spectroscopic redshifts spanning $4.0 < z < 11.44$ \citep{ref11}. The galaxies were retained only when their spectroscopic redshifts were considered robust. Physical properties, including stellar mass, UV dust attenuation, metallicity and dust mass, were derived through joint spectroscopic and photometric fitting with a spectrophotometric implementation of CIGALE \citep{ref11}. The adopted CIGALE configuration, parameter grid and priors are described in \cite{ref12}. This sample revealed a population of galaxies with extremely low dust attenuation, referred to as GELDAs, whose relative abundance increases above $z \simeq 8.8$ \citep{ref11}. A mock analysis (see Fig. 5 of \citealt{ref12}) indicates that our approach can reliably constrain dust masses down to approximately $M_{dust} = 10^5 M_\odot$. The line ratios in the observed spectra are strongly correlated with $M_{\rm dust}$ (correlation coefficient r = 0.874). The UV slope $\beta_{FUV}$ is also correlated with $M_{\rm dust}$ (correlation coefficient $r = 0.667$). This means that, first, the emission lines, and second, the continuum shape drive the estimation of the amount of energy transferred into the far-IR via the energy balance.

The second sample comprises 26 galaxies at $z > 8$ assembled to investigate the transition from SNe-produced dust to dust growth in the ISM \citep{ref12}. At the time of writing, this sample constituted an exhaustive compilation of known $z > 8$ galaxies satisfying the adopted selection criteria. Its selection is driven primarily by the NIRSpec spectroscopic flux limit. The analysis combines JWST-derived constraints on UV attenuation, stellar mass, metallicity and UV luminosity functions with physically-motivated attenuation models. The galaxy properties were derived using the same CIGALE implementation, parameter grid and priors as adopted for the first sample \citep{ref11}.

Potential AGN in the CEERS first sample were identified and discussed in \citet{ref11}. The nature of one of the objects in the second sample, GHZ2 at $z = 12.33$, remains debated \citep{ref19,ref20,ref21,ref22}, with both star-forming galaxy and AGN interpretations proposed. We include GHZ2 in the fiducial censored analysis and repeat both the change-point and bootstrap calculations after excluding it when testing the redshift break.

The third sample is the PIXIEDust compilation of ten spectroscopically confirmed galaxies at $8.2 < z < 14.2$ observed with ALMA and Northern Extended Millimetre Array (NOEMA) \citep{ref15}. The compilation combines nearly 200 h of submillimetre and millimetre observations and provides independent constraints on dust masses during the first approximately 700 Myr of cosmic history. We adopt the dust mass measurements and upper limits reported directly in \citet{ref15}, without recalculating them using the dust emission model from \citet{ref12}. In \citet{ref15} dust masses are estimated from the ALMA/NOEMA continuum fluxes assuming optically thin emission described by a single-temperature modified blackbody, with $T_{\rm dust}=50$ K, emissivity index $\beta_{\rm dust}=2$, and a fixed dust opacity normalization; the calculation also accounts for CMB heating and the reduced contrast of dust emission against the CMB at high redshift.

For non-detections, these assumptions are used to convert the measured flux-density limits into \(3\sigma\) upper limits on \(M_{\rm dust}\), corrected for gravitational lensing where appropriate.

We combine the two JWST-based samples with the PIXIEDust ALMA/NOEMA compilation and interpret the resulting trends using a framework inspired (see Fig.~\ref{fig:architecture} and Sect.~\ref{subsec:compact_model}) by \citet{ref12}. The measurements, detections and upper limits used in the present analysis are reported in the source data tables accompanying this paper.

Together, these three datasets provide complementary constraints on UV attenuation, dust mass and the transition from stellar dust production to ISM grain growth during cosmic dawn.

\subsection{Determination of the redshift break}
\label{subsec:zbreak_determination}

We quantified the redshift dependence of the dust observables using a censored single change-point analysis, applied independently to $\log_{10}(M_{\rm dust} / M_\odot)$, $\log_{10}(A_{\rm FUV})$, and $\log_{10}(M_{\rm dust} / M_{\rm star})$. Sources with finite redshifts, measurements and associated uncertainties were retained. Candidate redshift breaks were defined as the midpoints between consecutive observed redshifts and restricted to $5.0 < z_{\rm break} < 13.5$.

For a candidate break, the underlying population was described by two constant latent levels. A common intrinsic scatter, $\sigma_{\rm int}$, describes the real galaxy-to-galaxy dispersion around these population means. For each detected galaxy $i$, the likelihood was Gaussian with total variance $\sigma_i^2+\sigma_{\rm int}^2$, where $\sigma_i$ is the measurement uncertainty. For an upper limit $y_i^{\rm lim}$, the likelihood contribution was the Gaussian cumulative probability. Because the high-redshift sample is dominated by upper limits, the inferred latent population level may also depend on the adopted Gaussian form of the censored likelihood; we therefore interpret the change-point result as a descriptive population-level statistic rather than a model-independent reconstruction of the intrinsic distribution.

\begin{equation}
\mathcal{L}_i^{\rm UL} = \Phi\!\left[\frac{y_i^{\rm lim}-\mu_i}{\sqrt{\sigma_i^2+\sigma_{\rm int}^2}}\right],
\end{equation}

\noindent where $\mu_i$ is the latent population mean predicted on the corresponding side of the break and $\Phi$ is the standard normal cumulative distribution function. For each candidate redshift break, the two latent levels and intrinsic scatter were optimized, and the break maximizing the total censored likelihood was retained.

The analysis was performed both including and excluding GHZ2. The full-sample result is adopted as fiducial, while the calculation without GHZ2 tests sensitivity to this possible AGN. No physical model track, metallicity threshold, or prior centred on $z = 9$ enters the empirical break determination.

Sampling uncertainty in $z_{\rm break}$ was estimated using a non-parametric object bootstrap. Galaxies were resampled with replacement while retaining each object's redshift, measured value, uncertainty, and detection or upper-limit classification. The input catalogues were resampled separately to preserve their original sample sizes. We performed 2000 bootstrap realizations for the fiducial and no-GHZ2 samples. The quoted 68 \% and 95 \% intervals correspond to the $16^{\rm th} - 84^{\rm th}$ and $2.5^{\rm th} - 97.5^{\rm th}$ percentiles of the resulting redshift break distributions (see Sect.~\ref{subsec:zbreak}).

We additionally compared the one-break model with a no-break model using the same censored likelihood. The no-break model contains one latent population level and an intrinsic scatter, while the one-break model contains two levels, an intrinsic scatter, and the break position. Model preference was assessed using the Akaike and Bayesian information criteria \citep{ref25,ref26}, with $\Delta_{AIC}$ and $\Delta_{BIC}$ defined as the information criterion of the no-break model minus that of the one-break model, so that positive values favour a break. For completeness, we report both AIC and BIC differences, but use $\Delta_{BIC}$ as the primary descriptive model comparison statistic. 

The full sample values are $\Delta_{BIC}$ = 8.66, -4.75, and 14.87 for $M_{\rm dust}$, $A_{\rm FUV}$, and $M_{\rm dust} / M_{\rm star}$, respectively. After excluding GHZ2, the corresponding values are 11.71, -2.38, and 14.60. The full AIC, BIC, and likelihood comparison is reported in Table~\ref{tab:breaks}. The results will be discussed in Sect.~\ref{subsec:zbreak}.

\begin{table}[htbp]
\caption{Censored one-break versus no-break model comparison.
Differences are defined as the information criterion of the no-break model minus that of the one-break model, so positive values favour a break. The dust mass and dust-to-stellar mass ratio strongly favour a one-break description, whereas $A_{\rm FUV}$ does not independently require a break. The result is insensitive to the inclusion of GHZ2. Because the change-point
position is undefined under the no-break hypothesis, $2\Delta\ln L = 2 [\ln(L_{\rm break}) - \ln(L_{\rm nobreak})]$ is reported
descriptively and is not converted to a standard $\chi^2$ significance.}
\label{tab:breaks}
\centering
\setlength{\tabcolsep}{3pt}
\begin{tabular}{@{}lcccc@{}}
\hline\hline
Sample / observable & $z_{\rm break}$ & $2\Delta\ln L$ & $\Delta_{\rm AIC}$ & $\Delta_{\rm BIC}$ \\
\hline
Full: $M_{\rm dust}$ & 8.878 & 19.24 & 15.24 & 8.66 \\
Full: $M_{\rm dust}/M_{\rm star}$ & 8.878 & 25.46 & 21.46 & 14.87 \\
Full: $A_{\rm FUV}$ & 8.789 & 5.83 & 1.83 & -4.75 \\
No GHZ2: $M_{\rm dust}$ & 8.789 & 22.28 & 18.28 & 11.71 \\
No GHZ2: $M_{\rm dust}/M_{\rm star}$ & 8.789 & 25.18 & 21.18 & 14.60 \\
No GHZ2: $A_{\rm FUV}$ & 8.878 & 8.20 & 4.20 & -2.38 \\
\hline
\end{tabular}
\end{table}

\subsection{Benchmark dust evolution model}
\label{subsec:benchmark}

We adopt the model of \citet{ref18} as a benchmark because it provides a simple, widely used framework that includes the key competing processes of stellar dust production, SN destruction, and metallicity-dependent ISM grain growth. However, many alternative dust evolution models differ in their prescriptions for stellar yields, dense gas growth, destruction, and gas cycling, while retaining the same broad competition between stellar dust production and ISM grain growth. The model follows the coupled evolution of gas, stars, metals and dust and includes dust production by core collapse supernovae and asymptotic giant branch (AGB) stars, dust destruction, and metallicity-dependent grain growth through the accretion of gas-phase metals in the ISM. Full details of the chemical evolution equations, stellar yields, dust destruction prescription and grain growth formalism are given in \citet{ref18}.

We calculate model tracks for galaxies beginning their evolution at formation redshifts $z_{\rm form} =$ 25, 20 and 15. For each formation redshift, the model separately follows the dust mass contributed by core collapse SNe, AGB stars and ISM grain growth, together with the total dust mass. The resulting evolution of $M_{\rm dust}$ and $M_{\rm dust} / M_{\rm star}$ is compared with the observational sample in Sect.~\ref{subsec:dust_processes}.

At the earliest times, core collapse SNe provide the dominant source of newly condensed dust and seed the ISM with the grains required for subsequent growth. The contribution from AGB stars is delayed by the longer evolutionary timescales of their progenitors and remains negligible at $z > 4$ in the benchmark calculations. Dust formation during cosmic dawn is therefore governed primarily by SNe-condensed grains and, at later times, by grain growth in the ISM.

As chemical enrichment proceeds, the grain growth timescale decreases and the accretion of gas-phase metals onto pre-existing grains becomes increasingly efficient. The model consequently predicts a transition from a stellar source-dominated dust regime to one dominated by ISM grain growth.

The benchmark model therefore demonstrates that metallicity-dependent grain growth can generate the required qualitative transition in dust production, although its evolution is more gradual than the population change identified empirically around $z \simeq 8.8$.

All model ingredients and parameter values are adopted from \citet{ref18} unless stated otherwise.

\subsection{Compact high-redshift galaxy model}
\label{subsec:compact_model}

\begin{figure*}[t]
\centering
\includegraphics[width=\linewidth]{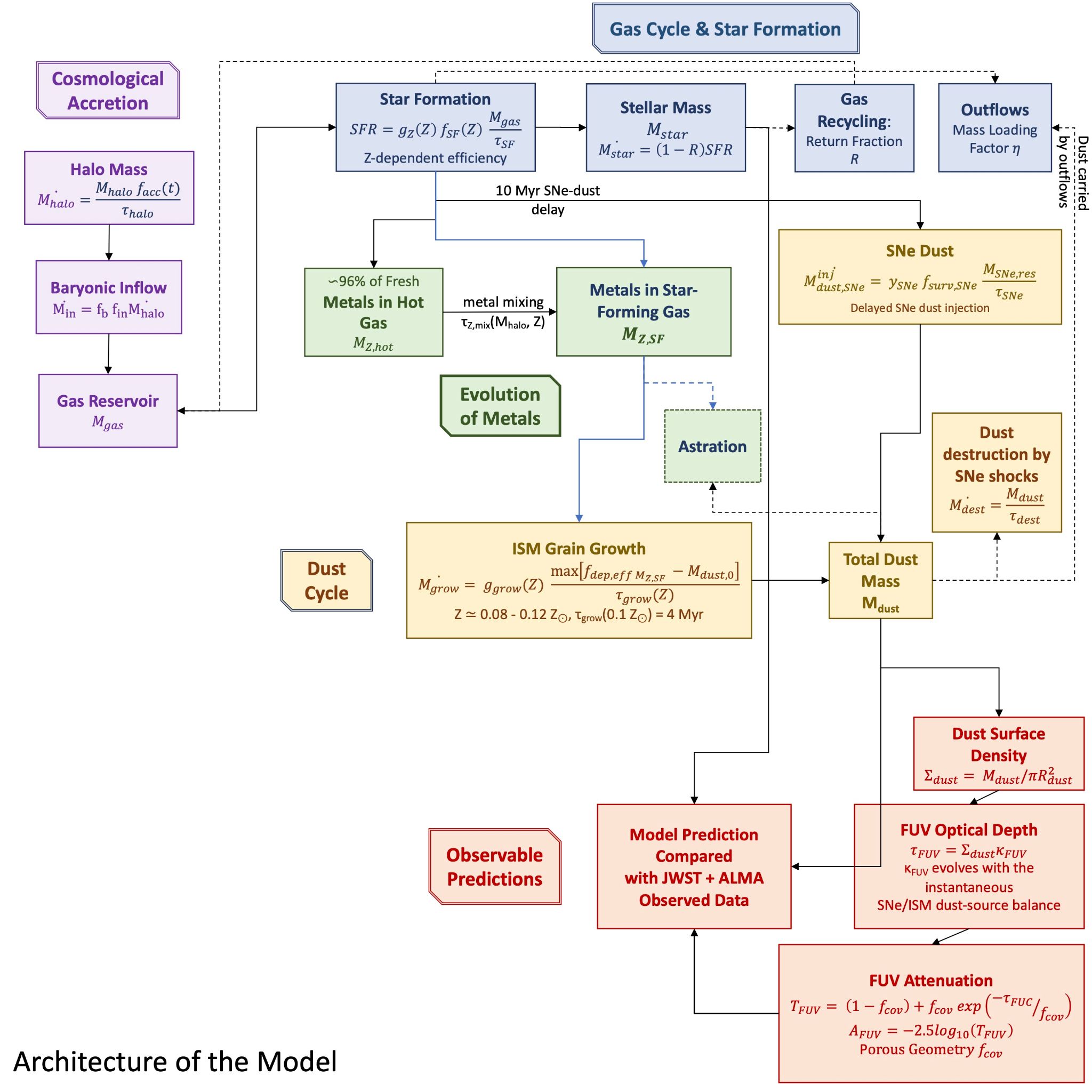}
\caption{\textbf{Architecture of the compact high-redshift galaxy model.} Schematic representation of the coupled evolution of halo mass, gas, stars, metals and dust in the fiducial model. Solid arrows indicate production, mass transfer or observable mappings, whereas dashed arrows indicate recycling or removal processes. Fresh stellar metals are deposited predominantly in a hot reservoir before mixing into the star-forming ISM. Dust is supplied by delayed SNe injection and metallicity-dependent ISM grain growth, and is removed by destruction, outflows and astration. The resulting dust mass and geometry determine the dust surface density, FUV optical depth and attenuation compared with the JWST and ALMA observations.}
\label{fig:architecture}
\end{figure*}

To interpret the population transition identified empirically, we use a one-zone chemical evolution model tailored to compact high-redshift galaxies (Fig.~\ref{fig:architecture}). The model follows the coupled evolution of gas, stars, metals and dust under halo-driven gas inflow, star formation, outflows, stellar metal and dust production, ISM grain growth and dust destruction. The chemical evolution is connected to UV attenuation through the evolving dust surface density, dust opacity and covering fraction. The empirical break redshift is determined independently of these model calculations; the model is subsequently compared with the observed redshift evolution of $M_{\rm dust}$, $M_{\rm dust} / M_{\rm star}$, and $A_{\rm FUV}$.

Halo growth is described through its specific accretion rate,

\begin{equation}
\frac{d\ \ln(M_{\rm halo})}{dt} = \ \frac{f_{\rm acc}(t)}{\tau_{\rm halo}},
\end{equation}

\noindent so that the absolute accretion rate scales with halo mass. The dimensionless factor $f_{\rm acc}(t)$ declines smoothly from unity at early times to a late-time floor $f_{\rm late}$. The parameter $t_{\rm decline}$ sets the midpoint of this decline, while $\Delta~t_{\rm decline}$ controls the temporal width of the decline where

\begin{equation}
f_{\rm acc}(t)\  = \ f_{\rm late}\  + \ (1 - f_{\rm late}\ )\left\lbrack 1 + \exp\left( \frac{t - t_{\rm decline}}{\Delta t_{\rm decline}} \right) \right\rbrack^{- 1}.
\end{equation}

The corresponding halo growth rate is

\begin{equation}
{\dot{M}}_{\rm halo} = \frac{M_{\rm halo}\, f_{\rm acc}(t)}{\tau_{\rm halo}}\,
\end{equation}

\noindent and the baryonic inflow rate is

\begin{equation}
{\dot{M}}_{in} = f_{b}\, f_{in}\,{\dot{M}}_{\rm halo}.
\end{equation}

In our model, we have adopted $\tau_{\rm halo} = 100$ Myr, $t_{\rm decline} = 600$ Myr,
$\Delta~t_{\rm decline} = 75$ Myr and $f_{\rm late} =$ 0.05. The accretion rate therefore declines smoothly at late times. No explicit transition in halo accretion is imposed at $z \simeq 9$.

Model parameters are adopted from the literature, held fixed, explored or calibrated, as indicated in Table~\ref{tab:parameters}.

We integrate the coupled equations from $z_{\rm form} =$ 25 over the redshift interval relevant to the observations, adopting a flat $\Lambda$CDM cosmology with $H_0 =$ 67.4 km\,s$^{-1}$ Mpc$^{-1}$, $\Omega_m =$ 0.315, and $\Omega_\Lambda =$ 0.685. Eleven initial halo masses are logarithmically spaced over $10^7 \le M_{\rm halo,0} / M_\odot \le 10^{11}$. 

For each halo, we make predictions for models with four sets of the following parameters: star formation efficiency, low-metallicity locking, effective surviving SNe dust, and dense-gas fraction ($f_{SF,post}$, $R_{lowZ}$, $f_{\rm SN,surv}$, $f_{dense}$ ; see Tab.~\ref{tab:parameters}). These four models sample the correlated variations in each input parameter, giving 44 model tracks in total. The coupled equations were integrated using an adaptive fifth-order Runge - Kutta method with an embedded fourth-order estimate for local error control. The internal integration timestep varied adaptively between model tracks and was limited to a maximum of 0.25 Myr for the fiducial calculations. For subsequent analysis and plotting, the numerical solution was evaluated on a common grid of 2600 equally spaced time points using the solver's Runge - Kutta interpolant. The fiducial calculations shown in the main figures do not include Population III star formation.

The four states are not varied independently but are paired to represent coherent physical regimes. We denote by $f_{\rm SF,post}$ the branch-dependent star-formation efficiency factor approached after the enrichment transition, by $R_{\rm lowZ}$ the effective stellar return fraction in the low-metallicity regime, by $f_{\rm dense}$ an effective fraction of the star-forming gas participating in efficient dense-ISM grain growth, and by $f_{\rm SN,surv}$ the scaling of the effective surviving SNe-condensed dust yield after reverse-shock and local ISM processing. The parameter $f_{\rm dense}$ is a phenomenological one-zone quantity rather than a fraction derived from an explicit gas-density threshold or density probability distribution.

In the model, $f_{\rm SF,post}$ is activated smoothly over $0.11 < Z/Z_{\odot} < 0.15$; this prescription represents a change in the effective star-formation state of the enriched ISM rather than an independently predicted dependence of star-formation efficiency on metallicity. From the most efficient to the least efficient post-transition star formation, the adopted combinations are ($f_{SF,post}$, $R_{lowZ}$, $f_{dense}$) = ($1, 0.65, 0.01$), ($0.3, 0.75, 0.03$), ($0.1, 0.84, 0.07$), and ($0.03, 0.90, 0.15$), with $f_{SN,surv}$ varied across four representative states, spanning an effective surviving SN-dust yield multiplier from $\sim 0.8$ to $\sim 3.7$, with intermediate values sampling progressively more efficient survival of SNe-condensed dust.
These combinations are intended to span plausible correlated variations rather than to constitute independent parameter measurements. These parameter combinations are phenomenological and are intended to span the range of physical conditions expected in rapidly evolving, metal-poor galaxies rather than to represent four uniquely calibrated populations. Their ranges are motivated by theoretical and numerical studies showing substantial variations in star-formation efficiency with metallicity and environment, stellar return fractions with stellar-population properties, the efficiency of grain growth with the fraction of gas in dense clouds, and the survival of SN-condensed dust through reverse shocks \citep{ref42,ref43,ref44,ref45,ref46,ref29,ref47}.

At very low metallicity, we impose a nearly common SN-dominated sequence in order to avoid generating four artificially distinct branches before the onset of efficient ISM grain growth. In particular, the low-metallicity return fraction is taken to be close to \(R_{\rm lowZ}=0.72\), with a branch-to-branch variation of only \(\pm0.01\). The full branch dependence is then recovered smoothly over $0.075 < Z/Z_\odot < 0.110$, approaching the adopted values $R_{\rm lowZ} =$ 0.65, 0.75, 0.84, and 0.90. The dense gas and SNe dust parameters are treated similarly, with their branch dependence compressed toward a common low-metallicity sequence. The SNe dust parameter varied here should be interpreted as an effective surviving SN-dust yield parameter rather than as the physical reverse-shock survival fraction $f_{\rm SN,surv}$. The four adopted states span approximately a factor of five in effective surviving SN-dust yield. In addition, we allow a $\pm0.30$ dex variation in this effective yield at the lowest metallicities, which fades smoothly over $0.055 < Z/Z_\odot < 0.090$. All return-fraction branches subsequently converge smoothly to the standard Population-II value $R = 0.50$ over $0.08 < Z/Z_\odot < 0.12$.

\subsubsection{Gas, stellar and metal evolution}
\label{sebsec:gas_stars_metals}

The gas and stellar masses evolve as

\begin{equation}
\frac{dM_{gas}}{dt} = {\dot{M}}_{in} - (1 - R)\, \mathrm{SFR} - {\dot{M}}_{out},
\end{equation}

\begin{equation}
\frac{dM_{star}}{dt} = (1 - R)\, \mathrm{SFR},
\end{equation}

\noindent where $R$ is the stellar return fraction and ${\dot{M}}_{out} = \eta SFR$ is the outflow rate.

Star formation is regulated by the available gas reservoir and by a metallicity-dependent gate $g_Z(Z)$, which represents suppressed Population II star formation in chemically pristine gas. Rather than adopting a single low-metallicity efficiency, we explore four paired physical states in which the fraction of gas effectively locked at low metallicity and the post-transition star formation efficiency vary coherently with the dense gas fraction and surviving SNe dust normalization. This allows us to test whether the predicted dust transition persists across different early enrichment histories rather than relying on one specific star-formation prescription.

The metallicity gate $g_Z$ controls how rapidly the model moves from its low-metallicity regime to the fully activated regime. At metallicities below $(1-w_Z) Z_{\rm trans}$, where $w_Z$ is the fractional width of the metallicity interval over which the gate transitions from $g_{\rm floor}$ to unity, the gate is fixed at its minimum value $g_{\rm floor}$. Between $(1-w_Z) Z_{\rm trans}$ and $Z_{\rm trans}$, it increases smoothly from $g_{floor}$ to 1 following the cubic smoothstep function $3u^2-2u^3$. This cubic smoothstep function is used to provide a continuous transition with zero slope at both boundaries. Above $Z_{\rm trans}$, the gate is fully open,
$g_Z = 1$. The normalized variable

\begin{equation}
u\  = \ \frac{Z\  - \ (1\  - \ w_{Z})\ Z_{\rm trans}}{w_{Z}\ Z_{\rm trans}}
\end{equation}

\noindent runs from 0 at the start of the transition interval to 1 at
$Z = Z_{trans}$.

\begin{equation}
\mathrm{SFR}(t) = \,\ g_{Z}(Z)\ f_{\rm SF}(Z)\,\frac{M_{\rm gas}}{\tau_{SF}}.
\end{equation}

The metallicity interval over which $f_{\rm SF}$ changes is chosen to follow the onset of efficient ISM grain growth near $Z \simeq 0.1 Z_\odot$. In the model, all tracks therefore share the same star-formation efficiency during the low-metallicity, SNe-dominated phase, preserving a common enrichment history. Their different long-term star-formation efficiencies are activated only after the ISM has crossed the grain-growth transition, over $0.11 < Z/Z_\odot < 0.15$. The precise limits of this interval are thus model choices intended to provide a smooth post-transition response rather than a separately predicted critical metallicity. The corresponding post-transition gas-depletion times are therefore approximately 100, 333, 1000 and 3333 Myr. The baseline star-formation timescale is $\tau_{\rm SF} = 100$ Myr.

Stellar metals are divided between a hot reservoir and the star-forming ISM. We denote their metal masses by $M_{Z,hot}$
and $M_{Z,SF}$, respectively, and define the metallicity relevant for nebular emission and grain growth as

\begin{equation}
Z\  = \ \frac{M_{Z,SF}}{M_{\rm gas}}.
\end{equation}

A fixed fraction $f_{Z,hot} =$ 0.96 of newly synthesized metals is initially deposited in the hot phase. The two reservoirs evolve according to

\begin{equation}
\frac{dM_{Z,hot}}{dt}\  = f_{Z,hot}\ y_{Z}\ SFR\  - \frac{M_{Z,hot}}{\tau_{Z,mix}}\ ,
\end{equation}

\noindent and

\begin{equation}
\begin{aligned}
\frac{dM_{Z,SF}}{dt}
&= (1-f_{Z,hot})\,y_Z\,SFR
 + \frac{M_{Z,hot}}{\tau_{Z,mix}} \\
&\quad - Z(1-R)\,SFR
 - Z\dot{M}_{out}.
\end{aligned}
\end{equation}

Before the transition, the hot-to-star-forming ISM mixing time depends on halo mass and spans approximately 40 - 1750 Myr, multiplied by the fixed normalization $f_{mix,pre} =$ 0.9607. The effective mixing time adopted here should not be interpreted as the local radiative cooling time of the hot gas, but rather as a characteristic recycling timescale for newly produced metals to return from a hot/outflowing reservoir to the star-forming ISM. Cosmological simulations predict a broad distribution of such recycling times. For example, \citet{Angles-Alcazar2017} find wind recycling times extending from $\gtrsim$ 10 Myr to a few Gyr, with median values of $\sim$ 100 - 350 Myr. Earlier numerical studies found characteristic recycling times of order 1 Gyr and, in some cases, a dependence on halo mass, although both the normalization and mass dependence vary substantially among simulations. We therefore regard the adopted range as a phenomenological description intended to encompass the uncertainty associated with gas transport, fountain cycling, re-accretion, and phase mixing, rather than as a prediction of the local cooling time. Once the star-forming ISM reaches $Z \simeq 0.10 - 0.13 Z_\odot$, the effective mixing time decreases smoothly toward 25 - 80 Myr. This metallicity-dependent recycling releases metals accumulated in the hot phase without introducing an explicit redshift switch.

\subsubsection{Dust production, growth and destruction}
\label{subsec:dust_models}

In the fiducial model, dust is tracked as the sum of dust produced by SNe and dust grown in the ISM. AGB dust production is neglected because its contribution is expected to be negligible at $z > 4$, consistent with the benchmark calculations described in Sect.~\ref{subsec:benchmark}. The total dust mass is

\begin{equation}
M_{dust} = M_{dust}^{SNe} + M_{dust}^{grow}.
\end{equation}

Dust injection by core collapse SNe is delayed relative to star formation by a short progenitor reservoir,

\begin{equation}
\frac{dM_{SNe,res}}{dt}\  = \ SFR\  - \frac{M_{SNe,res}}{\tau_{SNe}}\ ,
\end{equation}

\noindent with $\tau_{SNe} = 10$ Myr. The surviving SNe-condensed dust injection rate is

\begin{equation}
{\dot{M}}_{dust,\ SNe}^{inj} = \, y_{dust,\ SNe}^{}\ f_{SNe,\ surv}\ \ \frac{M_{SNe,res}}{\tau_{SNe}},
\end{equation}

\noindent where $f_{SN,surv}$ represents the effective survival of newly condensed grains after reverse shock and local ISM processing. It is varied as part of four predefined correlated physical states that also vary the post-transition star-formation efficiency, low-metallicity gas return, and dense-gas fraction.

ISM grain growth is treated as relaxation toward a finite equilibrium dust mass set by the amount of metals available in the star-forming ISM. We define

\begin{equation}
M_{\rm dust,\ eq} = \ f_{\rm dep,\ eff}\ M_{Z,\ \mathrm{SF}},
\end{equation}

\noindent where \(M_{\rm dust,eq}\) is the maximum dust mass allowed by the current metal reservoir in the star-forming ISM for the adopted depletion fraction \(f_{\rm dep,eff}\). $M_{Z,\rm SF}$ is the metal mass in the star-forming ISM, and $f_{\rm dep,eff}$ is the effective fraction of those metals that can be depleted onto dust grains. Here, $M_{\rm dust,eq}$ is not a separate dust reservoir, but the instantaneous depletion-limited target set by the current metal content of the star-forming ISM.

The fiducial depletion target is \(f_{\rm dep,max}=0.82\), with modest state-dependent variations. Grain growth is then implemented as

$$
\left(\frac{{\rm d}M_{\rm dust}}{{\rm d}t}\right)_{\rm grow}
=
g_{\rm grow}(Z)\,
\frac{\max\!\left[f_{\rm dep,eff}M_{Z,\rm SF}-M_{\rm d},\,0\right]}
{\tau_{\rm grow}(Z,f_{\rm dense})}.
$$

The global grain-growth gate \(g_{\rm grow}\) rises smoothly toward unity over \(0.08<Z/Z_\odot<0.12\), while below this interval only the dense-gas fraction contributes efficiently. 

We adopt a grain growth timescale of 4 Myr at $Z = 0.1Z_\odot$, with $\tau_{grow} \propto Z^{-1}$. The inverse metallicity dependence follows standard grain growth models, while the 4 Myr normalization is chosen to represent efficient growth in the dense ISM of compact high-redshift galaxies. The effective grain-accretion timescale depends on both metallicity and the dense-gas fraction,
\begin{equation}
\tau_{\rm grow}(Z,f_{\rm dense})
=
\tau_{\rm grow,crit}
\left(\frac{Z}{Z_{\rm crit}}\right)^{-1}
\mathcal{D}(f_{\rm dense}),
\end{equation}
where
\begin{equation}
\mathcal{D}(f_{\rm dense})
=
\left(
\frac{f_{\rm dense}}{f_{\rm dense,ref}}
\right)^{-1}.
\end{equation}
We adopt
$\tau_{\rm grow,crit}=4$ Myr, $Z_{\rm crit}=0.1\,Z_\odot$, and $f_{\rm dense,ref}=0.10$. Thus, branches with larger dense-gas fractions undergo more rapid grain growth. In the numerical implementation, $0.2\leq\mathcal{D}\leq8$, and the resulting accretion timescale is limited to $0.5\leq\tau_{\rm grow}/{\rm Myr}\leq5000$.

A small amount of sub-threshold growth is allowed in the dense gas component, but widespread ISM grain growth occurs only as the system enters the $0.08 - 0.12 Z_\odot$ regime.

Dust is removed through destruction, outflows and astration,

\begin{equation}
{\dot{M}}_{dust,loss} = \frac{M_{dust}}{\tau_{dest}} + \frac{M_{dust}}{M_{gas}}\,{\dot{M}}_{out}\  + \ \frac{M_{dust}}{M_{gas}}\ (1 - R)\ SFR,
\end{equation}

\noindent where $\dot{M}_{dust,loss}$ is the total dust-loss rate from destruction, removal in outflows and incorporation into long-lived stars and stellar remnants. $\tau_{dest}$ depends on the star formation rate (SFR). The SFR dependent destruction timescale is additionally rescaled by the environmental dust state, allowing the four paired physical states to represent correlated variations in grain survival and destruction. The destruction-state factor is allowed to vary over $0.08\leq f_{\rm dest,state}\leq5$, spanning nearly two orders of magnitude in the effective destruction timescale. This parameter describes uncertainty in the subsequent destruction of dust already present in the ISM and is distinct from the survival of newly condensed SN dust, which enters separately through the SN-dust injection term.

The outflow rate is assumed to scale with the star-formation rate according to

\begin{equation}
{\dot{M}}_{out} = \eta\ M_{halo}\ SFR,
\end{equation}

\noindent where the mass-loading factor is

\begin{equation}
\eta\ M_{halo}\  = \ clip\ \left\lbrack \left( \left( \frac{M_{halo}}{10^{10}M_{\odot}} \right)^{- 1/2} \right),\ 0.02,\ 10 \right\rbrack
\end{equation}

Here $M_{halo}$ is the instantaneous halo mass. The normalization gives $\eta =$ 1 at $M_{halo} = 10^{10} M_\odot$ and lower-mass haloes therefore experience stronger outflows, while the clipping limits prevent unrealistically large or small loading factors.

\subsubsection{Connection to ultraviolet attenuation}
\label{subsec:dust_attenuation}

To connect the chemical-evolution model to UV attenuation, we first compute the characteristic galaxy size using the high-redshift size scaling adopted in Table~\ref{tab:parameters}, motivated by \citet{ref38} and \citet{ref39},

\begin{equation}
R_e = R_{e,0}
\left(\frac{M_{\rm star}}{10^9\,M_\odot}\right)^\alpha
\left(\frac{1+z}{10}\right)^{-\beta},
\end{equation}

\noindent with $R_{e,0} =$ 0.8 kpc, $\alpha =$ 0.20, and $\beta =$ 0.30. The effective dust radius is allowed to differ from the stellar effective radius according to

$$
R_{\rm d}
=
R_{\rm e}\,f_R(\chi_{\rm geom}),
$$

\noindent where

$$
f_R(\chi_{\rm geom})
=
{\rm clip}
\left[
\chi_{\rm geom}^{-0.30},
\,0.50,\,
1.55
\right].
$$

Thus, larger geometry state values correspond to more compact dust distributions. The geometry state evolves from a common value $\chi_{\rm geom} =$ 1 in the SNe-dominated regime toward the branch-specific dust state $\chi_{\rm dust}$ as the ISM-grown dust contribution becomes important,

$$
\chi_{\rm geom} = \chi_{\rm dust}^{\,f_{\rm ISM,opt}}.
$$

In the SNe-dominated regime, the different physical states have similar effective dust radii and covering fractions, and their full range of compactness and covering fraction develops smoothly as ISM-grown dust becomes important. The attenuation calculation adopts an effective partial-covering geometry rather than a strictly uniform foreground screen. The mean dust surface density is approximated as

$$
\Sigma_{\rm dust} = \frac{M_{\rm dust}}{\pi R_{\rm dust}^{2}}.
$$

\noindent and the corresponding FUV optical depth is

\begin{equation}
\tau_{FUV} = \kappa_{FUV}\ \Sigma_{dust}.   
\end{equation}

The emergent FUV transmission is then

\begin{equation}
T_{FUV} = (1 - f_{cov}) + f_{cov}\, exp( - \frac{\tau_{FUV}}{f_{cov}}),
\end{equation}

Thus, a fraction \(1-f_{\rm cov}\) of the UV emission can escape through effectively low-column or porous sightlines, while the covered fraction experiences the effective dust column. This prescription is intended as a phenomenological representation of unresolved ISM porosity rather than a detailed model of the full column-density distribution expected from supersonic turbulence or the cavities and superbubbles produced by clustered stellar feedback.

Since the observed FUV attenuation is primarily sensitive to dust along sightlines toward young star-forming regions, we associate the effective FUV opacity with the local, contemporaneous dust-processing regime rather than with the mass-weighted composition of the entire surviving dust reservoir.

We first define the instantaneous fraction of the total dust-production
rate supplied by ISM grain growth as
\begin{equation}
f_{\rm source,raw}
=
\frac{\dot{M}_{\rm dust,grow}}
{\dot{M}_{\rm dust,grow}+\dot{M}_{\rm dust,SN}}.
\end{equation}
The source fraction entering the opacity transition is then
metallicity-gated,
\begin{equation}
f_{\rm source}(Z)
=
f_{\rm source,raw}\,G_Z(Z),
\end{equation}
where
\begin{equation}
G_Z(Z)=
\begin{cases}
0, & Z\leq0.075\,Z_\odot,\\
u^2(3-2u), & 0.075\,Z_\odot<Z<0.100\,Z_\odot,\\
1, & Z\geq0.100\,Z_\odot,
\end{cases}
\end{equation}
with
\begin{equation}
u=
\frac{Z-0.075\,Z_\odot}{0.025\,Z_\odot}.
\end{equation}

The transition from SN-like to ISM-like optical properties is represented by
\begin{equation}
f_{\rm ISM,opt}
=
\frac{f_{\rm source}^{\,q}}
{f_{\rm source}^{\,q}+f_{\rm source,50}^{\,q}},
\end{equation}

\noindent where $f_{\rm source,50} =$ 0.9268 is the source fraction at which the transition is half complete and $q =$ 8 controls its sharpness. The exponent $q$ is a phenomenological response parameter rather than a physical power-law index. The effective opacity is then

\begin{equation}
\begin{aligned}
\log(\kappa_{\rm FUV}) 
&= (1-f_{\rm ISM,opt})\log(\kappa_{\rm FUV,SNe}) \\ 
&\quad
+f_{\rm ISM,opt}\log(\kappa_{\rm FUV,ISM}).
\end{aligned}
\end{equation}

\noindent with $\kappa_{FUV,SN} = 10^3$ and $\kappa_{FUV,ISM} = 10^5$ cm$^2$\,g$^{-1}$ (derived from the analysis carried out in \citet{ref12}. The dust geometry and covering fraction evolve coherently with the same source transition, such that the different physical states remain relatively similar in the SNe-dominated regime and develop their full range of compactness once ISM grain growth becomes important. The optical depth remains

\begin{equation}
\tau_{FUV} = \kappa_{FUV}\ \Sigma_{dust}.   
\end{equation}

The baseline covering fraction increases with stellar mass according to

\begin{equation}
f_{cov,\mathrm{base}} = f_0 + (f_1-f_0) \left[1-\exp\left(-\frac{M_{star}}{M_0}\right) \right].
\end{equation}

\noindent where $f_0 =$ 0.50, $f_1 =$ 0.9995 and $M_0 = 3 \times 10^8 M_\odot$ \citep{ref12}. The uncovered fraction is then modified by the same environmental geometry state that controls the effective dust radius, so that the branches remain geometrically similar in the SNe-dominated regime and diverge as ISM processing becomes important.

To account for a porous ISM, we introduce a covering fraction f$_{cov}$, yielding a transmission (Eq. 12 in \citealt{ref27}):

\begin{equation}
T_{FUV} = (1 - f_{cov}) + f_{cov}\, exp( - \frac{\tau_{FUV}}{f_{cov}}),
\end{equation}

and attenuation 
\begin{equation}
A_{FUV} = - 2.5{\ log}_{10}(T_{FUV}). 
\end{equation}

The coupled equations are integrated from $z_{form} \approx 25$ to $z \approx 4$ over a range of halo masses, producing predictions for metallicity, dust mass, dust-to-stellar mass ratio, and UV attenuation that are directly compared with the JWST and ALMA observations.

\subsubsection{Population III extension}
\label{subsec:PopIII_extension}

The fiducial model used throughout the main analysis does not include Population III star formation. To test whether the inferred dust transition depends on the earliest enrichment history, we repeat the calculation with two Population III star-formation histories while leaving all parameters of the compact-galaxy model fixed. The first represents a continuous early Population III episode, whereas the second consists of five short Population III bursts. These calculations are robustness tests only and are not used to calibrate the model.

In the Population III extensions, the total star-formation rate is

\begin{equation}
SFR(t) = \,{SFR}_{III}(t)\  + \left\lbrack 1 - w_{III}(t) \right\rbrack\ g_{Z}(Z)\ f_{SF}(Z)\,\frac{M_{gas}}{\tau_{SF}},
\end{equation}

\noindent where $w_{III}$(t) describes the declining relative contribution of Population III enrichment. The delayed SNe-dust source receives separate Population III and Population II contributions, weighted by the evolving Population III fraction, while grain growth, dust destruction, metal recycling, geometry and the opacity prescription are otherwise identical to those of the fiducial model.

The effective stellar metal yield is

\begin{equation}
\, y_{Z}^{}\,\  = \ w_{III}\, y_{Z}^{III}\,\  + \ (1 - \ w_{III}\,)\ y_{Z}^{II}.
\end{equation}

For the Pop III component we adopt $R_{\rm III} = 0.60$, $y_{Z,\rm III} = 0.030$, and $y_{\rm D,SN,III} = 3\times10^{-4}$. The adopted return fraction and metal yield are representative IMF-averaged values motivated by zero metallicity stellar evolution and nucleosynthesis calculations, which predict substantial mass return and metal production but with strong dependence on progenitor mass, explosion energy, and fallback (e.g. \citealt{ref49, ref50}). The Pop III SNe dust yield is motivated by theoretical calculations of dust condensation in metal-free SN ejecta (e.g. \citealt{ref51}); we treat the adopted value as an effective surviving dust yield rather than as a uniquely predicted condensation efficiency. The Population III contribution declines smoothly according to

\begin{equation}
w_{\rm III} = \frac{1}{2} \left(1-\tanh\left(\frac{t-t_{\rm mid}} {\tau_{\rm PopIII,trans}}\right) \right).
\end{equation}

where $\tau_{PopIII,trans} = 10$ Myr. For the continuous case, $t_{mid} = t_{PopIII,end} + \tau_{PopIII,trans}$ whereas for the periodic case $t_{mid} = t_{last} + \tau_{PopIII,trans}$.

In the continuous realization, Population III star formation proceeds at $2\times 10^{-3} M_\odot$ yr$^{-1}$ from t = 0 to 10 Myr. In the periodic realization, five Gaussian bursts are distributed between t = 0 and 100 Myr, each with a peak SFR of
$2 \times 10^{-3} M_\odot$ yr$^{-1}$ and width 0.2 Myr.

The three calculations produce substantially different metallicity histories at the earliest epochs but nearly identical dust-transition epochs. The continuous Population III model gives median breaks $z = 9.148, 9.185$, and $9.148$ in $M_{\rm dust}, A_{\rm FUV}$ and $M_{\rm dust} / M_{\rm star}$, respectively, while the periodic model gives 9.110, 9.189, and 9.110. These differences are small compared with the observational uncertainty on the empirical break and demonstrate that the transition near $z \simeq 9$ does not require Population III enrichment.

\subsubsection{Parameter calibration and adopted model family}
\label{param_calib}

The compact-galaxy model tracks shown in this paper are generated from a frozen 44-track model family, with no parameter optimization performed when producing the final figures. During model development, two global response parameters were calibrated once and then fixed. 

The pre-transition mixing normalization, $f_{mix,pre} = $ 0.9607, sets the physical enrichment clock and was determined from the transition timing in $M_{\rm dust}$ and $M_{\rm dust} / M_{\rm star}$. We define $f_{\rm source,50}$ as the instantaneous fraction of the total dust-production rate contributed by ISM grain growth at which the transition from SN-like to ISM-like FUV opacity reaches its midpoint. The opacity source-balance midpoint was then calibrated to $f_{\rm source,50} =$ 0.9268 from the attenuation response, without modifying the gas, stellar, metal, or dust-mass histories. The empirical censored change-point analysis described above is performed separately and does not impose a redshift transition in the differential equations.

\section{Results}
\label{sec:results}

\subsection{Observational evidence for a dust transition near $z\simeq9$}
\label{subsec:zbreak}

Independent JWST and ALMA measurements show a marked change in galaxy dust properties around $z \simeq 9$. We quantify this behaviour with a censored change-point analysis in which detections enter through their measurement likelihoods and upper limits through one-sided likelihoods. The break position is determined directly from the data and is fitted independently for $M_{\rm dust}$, $A_{\rm FUV}$, and $M_{\rm dust} / M_{\rm star}$.
No model track or prior on the transition redshift enters this calculation.

\begin{figure*}[t]
\centering
\includegraphics[width=\linewidth]{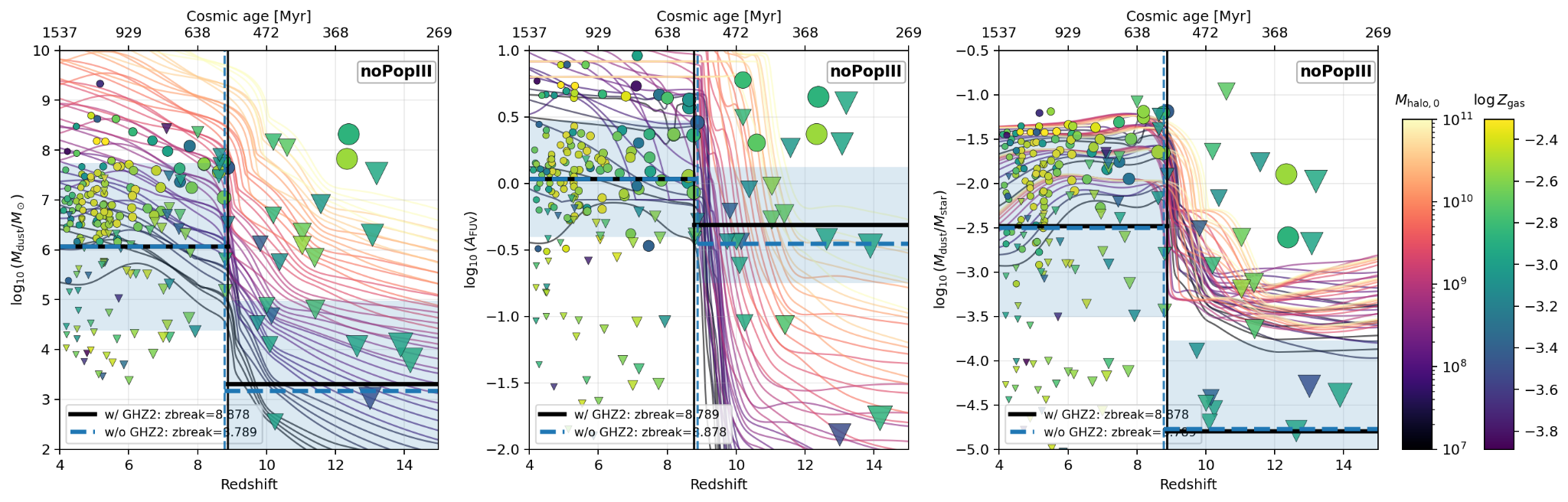}
\caption{\textbf{Evolution of dust properties across cosmic dawn.} Dust mass $M_{\rm dust}$ (left), FUV attenuation $A_{\rm FUV}$ (centre), and dust-to-stellar mass ratio $M_{\rm dust}/M_{\rm star}$ (right) are shown as a function of redshift. Circles indicate detections and downward triangles upper limits. Symbol colour indicates gas-phase metallicity. Curves show the fiducial model tracks without Population III enrichment, colour-coded by halo mass; these tracks are not used in determining the empirical break. Solid black lines show the maximum-likelihood censored one-break description of the full sample, and blue dashed lines show the corresponding result after removing GHZ2. The light blue regions show the fitted 68\% population ranges on either side of the full sample break and are not confidence intervals on $z_{\rm break}$. The preferred change point lies at $z\simeq8.8$ for all three observables and is nearly unchanged when GHZ2 is excluded. The two detections at $z>z_{\rm break}$ are the possible AGN GHZ2 at $z = 12.33$ discussed in Sect.~\ref{subsec:zbreak_determination}, and the lensed galaxy UNCOVER38766 at $z = 12.39$, for which the morphology could suggest an ongoing merger \citep{ref48}. If these alternate natures are confirmed, our model would not provide us with a valid representation.
}
\label{fig:dust-redshift}
\end{figure*}

For the full sample, the maximum likelihood break occurs at $z_{\rm break} =$ 8.878 for $M_{\rm dust}$, 8.789 for $A_{\rm FUV}$, and 8.878 for $M_{\rm dust} / M_{\rm star}$ (Fig.~\ref{fig:dust-redshift}). As we saw early on, GHZ2 might contain an active galactic nucleus, but removing GHZ2 only changes these values to 8.789, 8.878, and 8.789, respectively. Thus the preferred transition redshift changes by only $\Delta z \simeq 0.09$, showing that the result is not driven by this individual high-redshift source.

\begin{figure}[t]
\centering
\includegraphics[width=\linewidth]{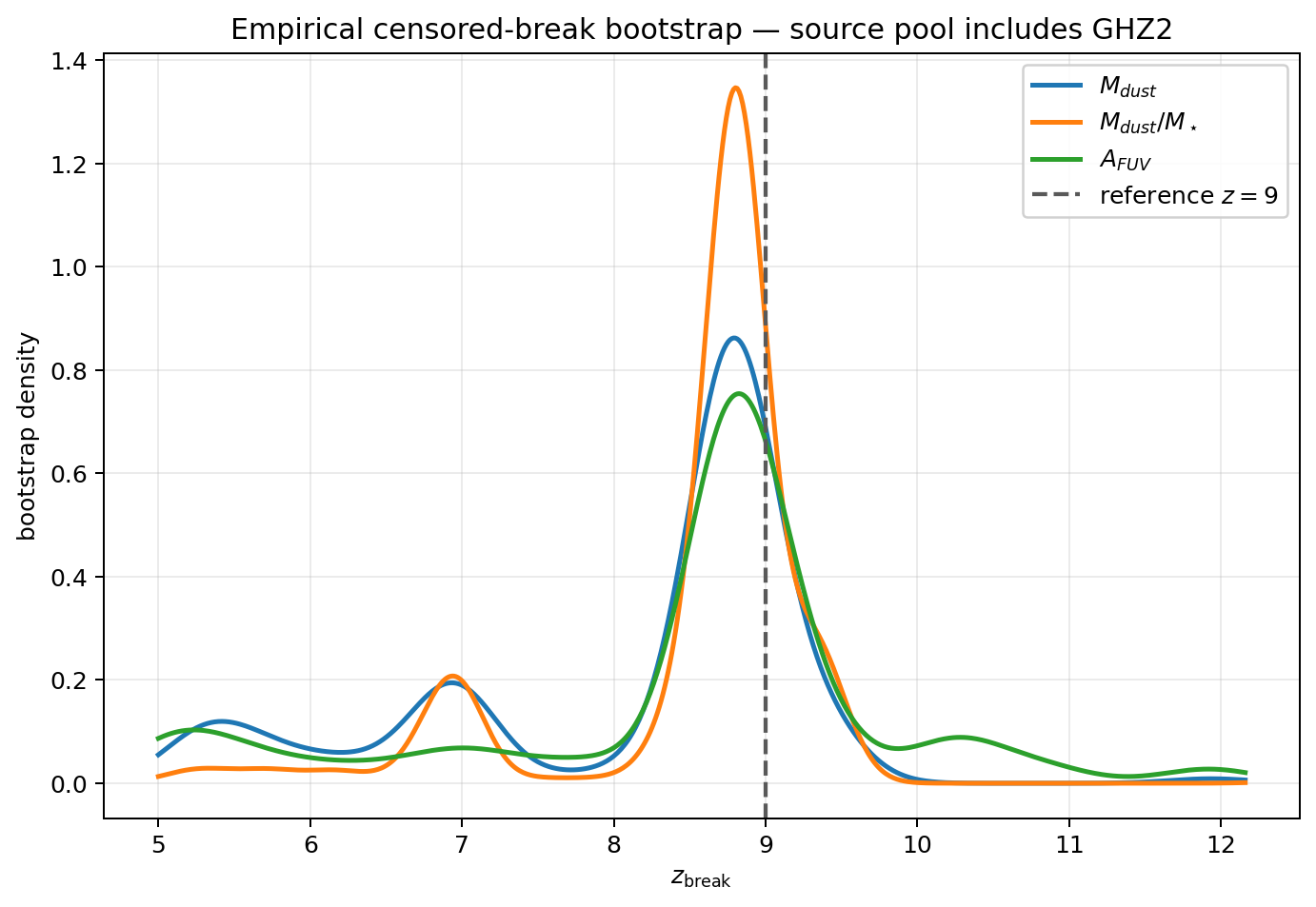}
\caption{\textbf{Bootstrap distribution of the empirical break redshift.} Kernel-density representations of the redshift break ($z_{\rm break}$) distributions obtained by non-parametrically resampling galaxies and repeating the complete censored change-point analysis. Blue, orange, and green curves correspond to $M_{\rm dust}$, $M_{\rm dust}/M_{\rm star}$, and $A_{\rm FUV}$, respectively. The dashed line at $z=9$ is shown for reference only and is not used as a prior or constraint. The density curves are visualizations of the bootstrap samples. Confidence intervals quoted in the text are calculated directly from their empirical quantiles.}
\label{fig:bootstrap}
\end{figure}

A non-parametric object bootstrap confirms a dominant solution near $z \simeq 8.8$ \citep{ref11}, although the redshift break distributions are non-Gaussian (Fig.~\ref{fig:bootstrap}). For 2000 bootstrap realizations of the full sample, the median and 68 \% intervals are $z_{\rm break} =$ 8.785 [6.900, 8.878] for $M_{\rm dust}$, 8.789 [8.405, 8.878] for $M_{\rm dust} / M_{star}$, and 8.789 [7.054, 9.308] for $A_{\rm FUV}$. The long tails, particularly for $M_{\rm dust}$, reflect the dominance of upper limits at the highest redshifts.

An explicit one-break versus no-break comparison using the same censored likelihood and the Bayesian information criterion (BIC) favours a break in $M_{\rm dust}$ and $M_{\rm dust} / M_{\rm star}$. 
For the full sample, the one-break model is preferred over the no-evolution model by $\Delta_{\rm BIC}=8.66$ for $M_{\rm dust}$ and $\Delta_{\rm BIC}=14.87$ for $M_{\rm dust}/M_{\rm star}$, where $\Delta_{\rm BIC}\equiv{\rm BIC}_{\rm null}-{\rm BIC}_{\rm break}$. These differences correspond approximately to Bayes factors of $76:1$ and $1.7\times10^3:1$, respectively, in favour of the one-break model. Under equal prior probabilities for the two models, these correspond to approximate posterior model probabilities of 98.7\% and 99.94\%. In contrast, $A_{\rm FUV}$ gives $\Delta{\rm BIC}=-4.75$, corresponding to odds of approximately $11:1$ in favour of the null model. The current attenuation measurements therefore do not independently require a break, although their maximum-likelihood change point lies at a similar epoch. Removing GHZ2 strengthens the result for the dust observables, yielding $\Delta_{BIC} =$ 11.71 and 14.60 for $M_{\rm dust}$ and $M_{\rm dust} / M_{\rm star}$, respectively, while $A_{\rm FUV}$ remains statistically consistent with no break $\Delta_{BIC} =$ -2.38. The redshift evolution of the three observables, together with the maximum-likelihood one-break descriptions, is shown in Fig.~\ref{fig:dust-redshift}.

Two detections at $z>z_{\rm break}$ lie above the fitted high-redshift population mean in Fig.~\ref{fig:dust-redshift}, illustrating the substantial intrinsic scatter and the limited number of detections that directly constrain the pre-transition population. One of them is GHZ2 that was discussed in Sect.~\ref{subsec:zbreak_determination} because of the possibility that this object might contain an AGN, in which case our model is not valid. The other one is UNCOVER38766 at $z = 12.39$. This object is lensed ($\mu \approx 1.5$). For this object, \citet{ref48} finds that an ongoing merger could also explain the observed clumpy morphology, in which case our model would also not be a valid representation. Conversely, upper limits occur on both sides of the fitted change point. The fitted levels therefore describe the censored population statistically and should not be interpreted as a separation of all individual galaxies into two non-overlapping populations. In particular, heterogeneous observational depths and target selection may contribute to the apparent difference between the two redshift regimes.

The sampling uncertainty in the inferred transition is shown in Fig.~\ref{fig:bootstrap}. The three bootstrap distributions have a dominant mode close to $z \simeq 8.8$, but are asymmetric and contain secondary solutions, particularly for $M_{\rm dust}$. This behaviour is expected because only a small number of dust detections are available above the preferred break, with most of the high-redshift constraints being upper limits. The relatively narrow dominant peak in $M_{\rm dust} / M_{\rm star}$, together with the one-break versus no-break model comparison, provides the strongest empirical evidence for a population transition.

\subsection{Identifying the main dust processes at play}
\label{subsec:dust_processes}

To identify which dust formation processes are at play, we first examine whether the observed transition in dust properties near $z \simeq 9$ can be reproduced by a benchmark dust evolution model \citep{ref18}. We use this model as a reference point before developing a framework better suited to the compact galaxies observed at $z \gtrsim 9$. The physical ingredients of the benchmark model are described in Sect.~\ref{sec:methodology}.

\begin{figure*}[t]
\centering
\includegraphics[width=\linewidth]{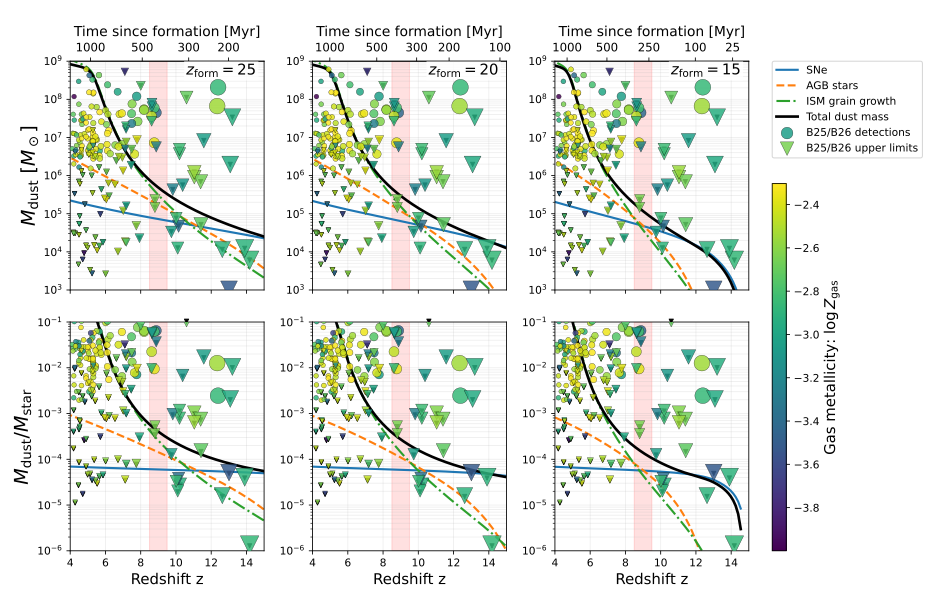}
\caption{\textbf{Dust evolution at cosmic dawn is dominated by SNe and ISM grain growth \citep{ref18}.} Dust mass $M_{\rm dust}$ (top) and dust-to-stellar mass ratio $M_{\rm dust}/M_{\rm star}$ (bottom) as a function of redshift for galaxies formed at $z_{\rm form}=25$, 20, and 15. Coloured curves show the contributions from core-collapse SNe, AGB stars, and ISM grain growth, while the black curve shows the total dust mass. Symbols denote the observational sample, with circles indicating detections and downward triangles indicating upper limits; symbol colour indicates gas-phase metallicity. The shaded region highlights $8.5 \lesssim z \lesssim 9 .5$ for visual reference around the preferred empirical change point. It is not a confidence interval on the transition width. AGB stars contribute negligibly to the dust-mass budget at $z > 4$, indicating that early dust evolution is governed primarily by dust produced by SNe and subsequently by grain growth in the ISM. The benchmark model predicts a gradual transition over several units in redshift, whereas the empirical change-point analysis identifies the strongest population change near $z \simeq 8.8$, without directly constraining its intrinsic physical width.}
\label{fig:benchmark}
\end{figure*}

Fig.~\ref{fig:benchmark} shows the evolution of dust mass and dust-to-stellar mass ratio for galaxies formed at different epochs. AGB stars contribute negligibly to the dust mass budget at $z > 4$, indicating that dust evolution during cosmic dawn is governed primarily by dust produced by core collapse SNe and, subsequently, by grain growth in the ISM. The benchmark model predicts a gradual transition from stellar-source-dominated dust production to grain growth dominated evolution over several units in redshift.

The empirical change-point analysis does not directly measure the physical width of this transition, but it shows that the observed population changes most strongly around $z \simeq 8.8$. The benchmark model captures the essential sequence from SNe-condensed dust to ISM grain growth, but predicts a more gradual evolution than the pronounced population contrast observed across this epoch. This motivates a model tailored to the compact high-redshift galaxy population revealed by JWST, in which metal recycling, grain growth, dust opacity and geometry evolve self-consistently.

\subsection{Dust mass and ultraviolet attenuation across the transition}
\label{subsec:transition}

The model predicts that dust enrichment remains slow while galaxies are metal poor, but accelerates rapidly once a critical metallicity threshold is reached. This transition marks the onset of efficient grain growth in the ISM and produces a corresponding increase in UV attenuation.

\begin{figure*}[t]
\centering
\includegraphics[width=\linewidth]{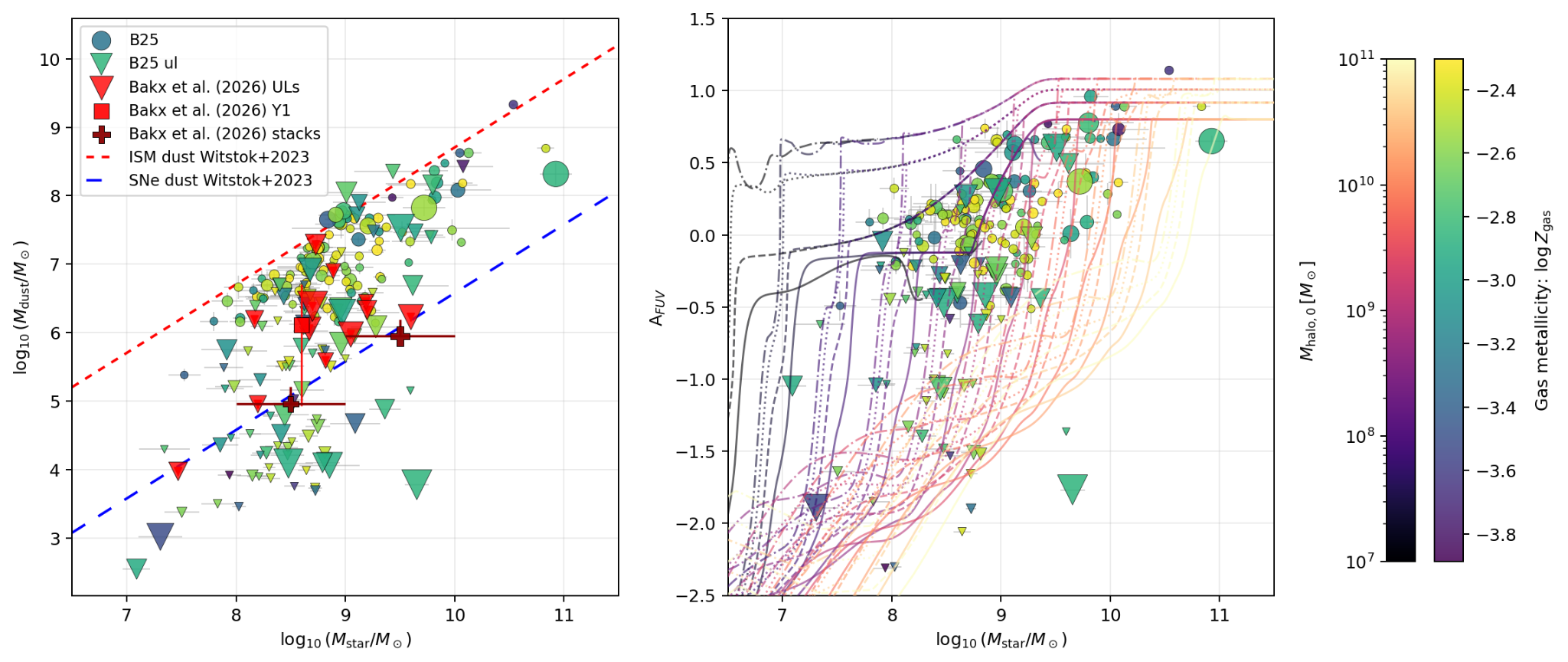}
\caption{\textbf{Dust mass and UV attenuation as a function of stellar mass.} Dust mass versus stellar mass (left) and FUV attenuation versus stellar mass (right) are compared with the fiducial model tracks. Circles denote detections and downward triangles upper limits; symbol size increases with redshift and colour indicates gas-phase metallicity. Model tracks span the adopted halo-mass and physical-state grid. Dashed lines in the left panel indicate the expected scaling relations for SNe-dominated and ISM-grain-growth-dominated dust regimes \citep{ref09}. The model reproduces the observed increase in dust mass and attenuation as galaxies evolve and become progressively enriched. The detection at $\log_{10}M_{\rm star} \sim 11$ is the same possible ongoing merger UNCOVER38766 in connection with Fig.~\ref{fig:dust-redshift}.}
\label{fig:dust-stellar}
\end{figure*}

Fig.~\ref{fig:dust-stellar} presents the full JWST and ALMA/NOEMA datasets in the $M_{\rm dust}$ - $M_{star}$ and $A_{\rm FUV}$ -
$M_{star}$, planes together with the fiducial model tracks spanning the adopted halo mass and physical state grid. A more detailed description is provided in Sect.~\ref{sec:methodology}. Galaxies with low dust masses and weak attenuation occupy the region expected for systems whose dust content is dominated by SNe-condensed grains (Fig.~\ref{fig:dust-redshift}). More chemically enriched galaxies populate the grain growth branch, where dust mass and attenuation increase rapidly once the characteristic metallicity regime is reached. The model reproduces both the observed $M_{\rm dust}$ - $M_{\rm star}$
relation and the emergence of strongly attenuated systems at later times.

\subsection{A model for compact high-redshift galaxies}
\label{subsec:hiz-model}

Motivated by the more gradual evolution predicted by the benchmark model, we have developed in Sect.~\ref{subsec:compact_model} a one-zone framework tailored to compact galaxies in the first billion years of cosmic history. As before \citep{ref18}, the central ingredient is metallicity-dependent grain growth, which becomes efficient after galaxies reach a characteristic enrichment regime $Z_{crit}$. The model additionally includes delayed recycling of freshly produced metals into the star forming ISM, compact galaxy sizes, evolving dust opacity and porous UV escape geometries \citep{ref12}. The physical clock controlling the dust mass evolution was calibrated using the $M_{\rm dust}$ and $M_{\rm dust} / M_{\rm star}$ transitions. The mapping between the instantaneous dust source balance and FUV opacity was calibrated separately using
$A_{\rm FUV}$, after which both response parameters were fixed. No parameters are re-optimized when generating the final model tracks.

The model follows the coupled evolution of gas, stars, metals and dust under halo-driven inflows, star-formation-driven outflows, SNe dust production and metallicity-dependent grain growth (Fig.~\ref{fig:architecture}).

A key feature of the model is that the UV opacity evolves together with the dust population. Dust surviving reverse-shock processing is expected to be dominated by relatively large grains and therefore exhibits a comparatively low FUV opacity. The onset of grain processing in the ISM increases both the dust mass and the UV absorption efficiency of the dust population, amplifying the observational signature of the transition.

Once galaxies reach the critical metallicity $Z_{crit}$ required for efficient grain growth, dust enrichment accelerates rapidly. In compact systems, this increase in dust mass is accompanied by higher dust surface densities and larger UV opacities, producing a sharp rise in attenuation. The model reproduces both the weak attenuation observed in the earliest galaxies and the rapid emergence of dust-rich systems at later times, providing a natural explanation for the observed transition near $z \simeq 9$.

\subsection{Robustness to Population III enrichment}
\label{subsec:PopIII}

Because some galaxies in the sample lie at $z > 10$, we test the sensitivity of the transition to Population III enrichment using two additional realizations of the final model: a continuous early Population III episode and a periodic Population III history. The same physical parameters and 44-track model family are used in all three cases; no parameter is recalibrated when Population III enrichment is introduced.

Population III enrichment strongly affects the earliest metallicity trajectories, particularly in the periodic case, where discrete enrichment episodes produce step-like increases in metallicity. Nevertheless, the epoch of the dust transition is remarkably stable. The median model redshift breaks in ($M_{dust}$, $A_{\rm FUV}$, $M_{dust} / M_{\rm star}$) are ($9.148, 9.182, 9.148$) without Population III enrichment, ($9.148, 9.185, 9.148$) for the continuous case, and ($9.110, 9.189, 9.110$) for the periodic case. Likewise, the median redshift break at which the star-forming ISM reaches $0.1 Z_\odot$ changes only from $z = 9.413$ without Population III enrichment to $9.413$ and $9.434$ in the continuous and periodic cases, respectively. Thus Population III stars alter the early enrichment path but do not set the transition near $z \simeq 9$, which instead remains associated with the onset of efficient ISM grain growth.

\begin{figure*}[t]
\centering
\includegraphics[width=\linewidth]{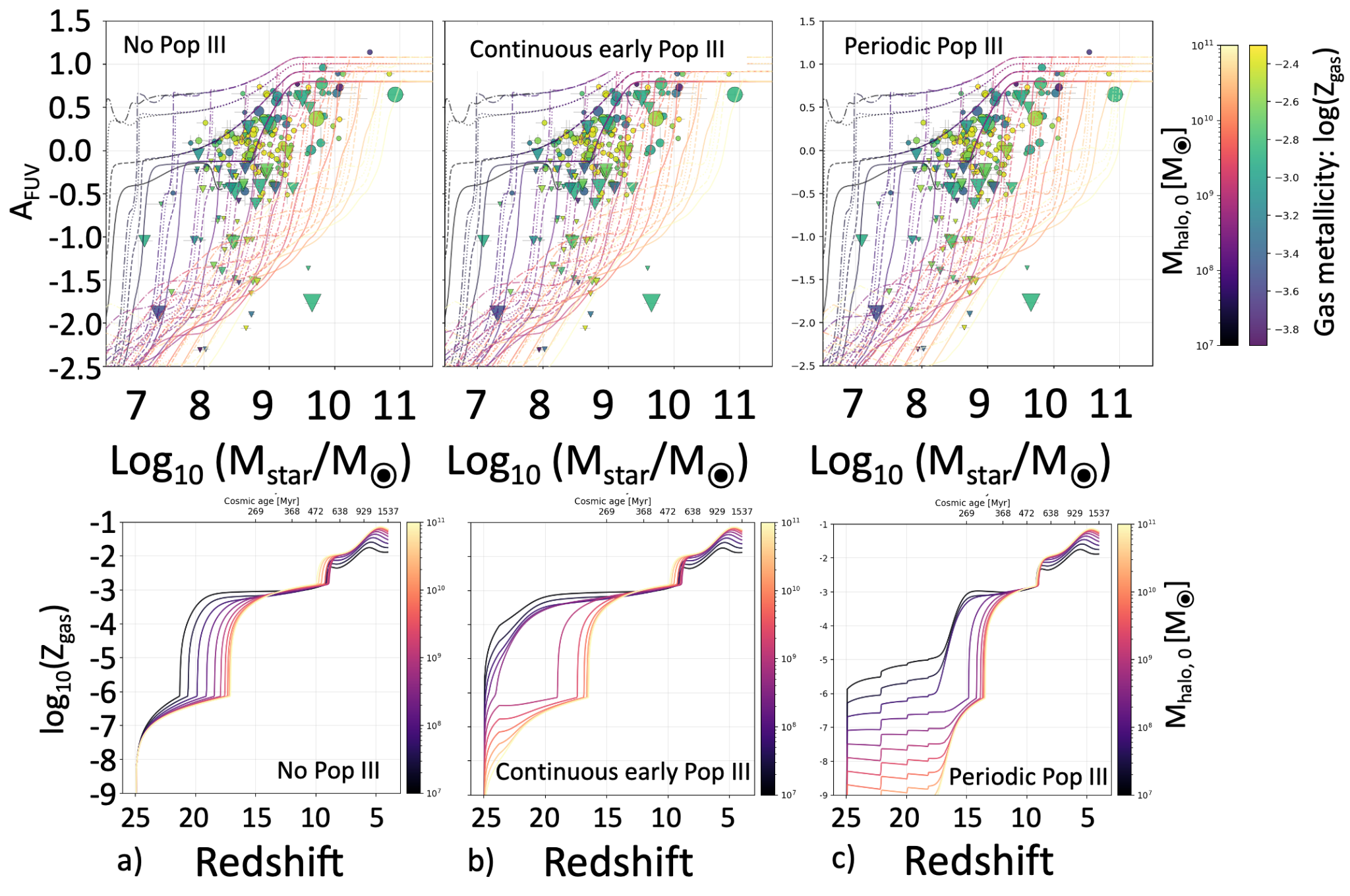}
\caption{\textbf{Population III enrichment modifies the early enrichment path but not the dust-transition epoch.} Top panels show $A_{\rm FUV}$ as a function of stellar mass for (a) the fiducial model without Population III enrichment, (b) a continuous early Population III episode, and (c) periodic Population III enrichment. Circles denote detections and downward triangles upper limits; colour indicates gas-phase metallicity and symbol size increases with redshift. Curves show the same 44-track model family in all three cases. Bottom panels show the corresponding evolution of the star-forming-ISM metallicity with redshift. The continuous Population III case modifies the earliest enrichment history, whereas periodic Population III enrichment produces pronounced step-like increases in metallicity. The median $0.1 Z_\odot$ crossing occurs at $z=9.413$, 9.413, and 9.434 for the no-Population-III, continuous, and periodic cases, respectively. The corresponding model breaks in ($M_{\rm dust}$, $A_{\rm FUV}$, $M_{\rm dust} /M_{\rm star}$) are ($9.148, 9.182, 9.148$), ($9.148, 9.185, 9.148$), and ($9.110, 9.189, 9.110$). Population III enrichment therefore changes the earliest chemical-enrichment path but is not responsible for the dust transition near $z \simeq 9$.}
\label{fig:popIII}
\end{figure*}

Fig.~\ref{fig:popIII} compares the fiducial, continuous and periodic Population-III realisations.

\section{Discussion}
\label{sec:Discussion}

The censored change-point analysis identifies a preferred transition near $z \simeq 8.8$ in the dust properties of early galaxies. The statistical evidence is strongest in $M_{\rm dust}$ and $M_{\rm dust} / M_{\rm star}$. The $A_{\rm FUV}$ attenuation measurements favour a change point at approximately the same epoch but do not independently require a break. The inferred location is essentially unchanged when GHZ2 is excluded.

Our modelling suggests that the observed population change is driven primarily by the onset of efficient ISM grain growth once galaxies reach a characteristic metallicity. At low metallicity, dust production is dominated by SNe-condensed grains and the dust content remains modest. As chemical enrichment proceeds, grain growth becomes increasingly efficient, increasing both dust mass and UV opacity. The correspondence between the empirical transition and the model behaviour is recovered across a range of halo masses and enrichment histories.

The role of Population III stars is primarily confined to the earliest enrichment history (Fig.~\ref{fig:popIII}). A continuous early Population III contribution changes the low-metallicity evolution while leaving the median $0.1 Z_\odot$ crossing and the observable dust breaks essentially unchanged. Periodic enrichment produces much more pronounced step-like metallicity evolution. However, even in this case the dust-mass and dust-to-stellar-mass breaks shift by only $\Delta z \simeq 0.04$, while the attenuation break changes by less than 0.01. The convergence of these different enrichment histories toward nearly the same transition epoch shows that the $z \simeq 9$ population change is controlled primarily by the metallicity-dependent recycling of metals and the onset of efficient ISM grain growth, rather than by the detailed history of the first stellar population. The observed transition should therefore not be interpreted as a direct signature of Population III stars. More detailed modelling exploring a broader range of Population III star formation histories and stellar properties will nevertheless be required.

The characteristic metallicity associated with the transition is model dependent and should not be interpreted as a direct empirical measurement. In the final model, widespread ISM grain growth is activated over $0.08 \lesssim Z/Z_\odot \lesssim 0.12$, while rapid recycling of metals from the hot reservoir into the star-forming ISM is activated over $0.10 \lesssim Z/Z_\odot \lesssim 0.13$. The observations constrain the epoch of the population change more directly than these metallicity intervals, which are parameters of the physical model used to interpret the transition.

An interesting by-product is that the transition may contribute to explaining the unexpectedly large abundance of UV-bright galaxies discovered by JWST at $z \gtrsim 9$ \citep[e.g.,][]{ref13,ref23,ref24}. If galaxies below the grain growth threshold contain little dust, a larger fraction of UV radiation can escape, boosting their apparent luminosities without requiring extreme star-formation efficiencies \citep{ref12}.

An alternative interpretation is that the apparent transition is enhanced by sample selection and censoring. Our censored analysis explicitly accounts for upper limits, and the preferred break location is robust to the removal of GHZ2. Moreover, $M_{\rm dust}$ and $M_{\rm dust} / M_{\rm star}$ favour a one-break model over a no-break model. The $A_{\rm FUV}$ measurements provide a consistent but statistically weaker indication of a transition at the same epoch. Larger samples, particularly additional dust detections at $z > 9$, will be required to determine the intrinsic width of the transition and quantify residual selection effects.

However, even though detecting the dust emission of a galaxy with $M_{\rm dust} \simeq 10^{6} M_\odot$ at $z \gtrsim 9$ is within reach of deep ALMA observations, detecting one with $M_{\rm dust} \simeq 10^{5} M_\odot$ remains excessively difficult with the present ALMA, without a strong coordinated effort of the community. However, this type of object and the associated major scientific question related to the first dust grains and their nature could benefit for the upgrade of ALMA presently under consideration. Independent constraints from JWST/MIRI Balmer-decrement and UV-slope measurements could help test whether the $z \gtrsim 9$ galaxies were indeed dust-poor and determine whether they contained sufficient dust reservoirs to seed the subsequent grain-growth phase.

\section{Conclusions}
\label{sec:conclusions}

The population transition favoured by the data may mark a critical phase in galaxy evolution. Once sufficient metals accumulate, the model predicts a rapid increase in both dust mass and UV opacity, transforming relatively transparent systems into dust-rich galaxies. The break around $z \simeq 9$ may therefore mark the emergence of dust as a major regulator of the radiative properties of galaxies and the population-scale transition from a SNe-dominated dust regime to one dominated by ISM grain growth. Population III enrichment can substantially modify the earliest buildup of metals and dust seeds, but the transition epoch itself is nearly unchanged and does not require Population III stars. 

Further progress will require both: a) future JWST and ALMA observations to test the present interpretation using larger samples at earlier epochs, improving our view on the buildup of metals, dust and opacity during cosmic dawn and b) developing a specific physical dust evolution and radiative-transfer modelling that could predict the observable signatures of competing scenarios, including threshold-like and smooth dust growth.

\begin{acknowledgements}
HSBA gratefully acknowledges support from Academia Sinica through grant AS-PD-1141-M01-2. DB, TD, and CA acknowledge support from the Centre National d'Etudes Spatiales (CNES). TB, CE, KK acknowledge support from the Knut and Alice Wallenberg Foundation (KAW 2020.0081).
\end{acknowledgements}

\clearpage
\onecolumn
\appendix

\section{ALMA observability: }Assuming dust temperature increases with redshift, as possibly suggested by recent observational and theoretical studies of compact high-redshift galaxies \citep{ref28,ref29}, and adopting post-reverse shock SNe dust opacities appropriate for large $\gtrsim 1 \mu$m surviving grains, consistent with predictions from fits to the JWST observations \citep{ref11,ref12}, a dust mass of $10^6 M_{\odot}$ at $z \simeq 10$ would produce a continuum flux of order
$20 - 30\ \mu$Jy at 1.2 mm and could be detectable in deep ALMA observations of a few hours under the adopted dust temperature and opacity assumptions.

In contrast, a dust mass of $10^5 M_\odot$ would produce a flux of only a few $\mu$Jy requiring substantially longer integrations and remaining challenging for current observations. The transition identified here therefore lies close to the practical sensitivity limit of present ALMA surveys, helping to explain the growing number of non-detections above $z \simeq 9$. Once galaxies reach the critical metallicity $Z_{crit}$ required for efficient grain growth, both dust mass and UV attenuation increase rapidly, producing the observed break around $z \simeq 9$.

\section{Supplementary model material}

\begin{longtable}{p{0.22\textwidth}p{0.12\textwidth}p{0.28\textwidth}p{0.32\textwidth}}
\caption{{\bf Main parameters controlling the compact high-redshift galaxy model.} The model follows 11 initial halo masses and four paired physical states, producing 44 evolutionary tracks. The paired states coherently vary the post-transition star formation efficiency, low-metallicity return fraction, surviving SNe dust contribution, and dense gas fraction. Efficient ISM grain growth is triggered by metallicity over $0.08 < Z/Z_\odot < 0.12$. No explicit redshift threshold is imposed. Two global response parameters, $f_{\rm mix,pre} =$ 0.9607 and $f_{\rm source,50} =$ 0.9268, are calibrated once from the transition timing and attenuation response and are subsequently held fixed when generating the final model family. Population III parameters apply only to extension tests; the fiducial calculation shown in the main analysis excludes Population III star formation.\label{tab:parameters}}\\
\hline\hline
Parameter & Symbol & Value & Source / status \\
\hline
\endfirsthead
\caption{Main parameters controlling the compact high-redshift galaxy model (continued).}\\
\hline\hline
Parameter & Symbol & Value & Source / status \\
\hline
\endhead
\hline
\endfoot
Star-formation timescale & $\tau_{\rm SF}$ & 100 Myr & This work \\
Return fraction (Pop III) & $R_{\rm PopIII}$ & 0.60 & \citet{ref49}; Pop III extension only, adopted \\
Return fraction (Pop II) & $R_{\rm PopII}$ & 0.50 & \citep{ref31} \\
Metal yield (Pop III) & $y_{Z,\rm PopIII}$ & 0.030 & \citet{ref49}; Pop III extension only, adopted \\
Metal yield (Pop II) & $y_{\rm Z,PopII}$ & 0.025 & \citep{ref31} \\
Dust yield  (SNe Pop III) & $y_{\rm D,SN,PopIII}$ & 3.0 $\times$ 10$^{-4}$ & \citep{ref51}, Pop III extension only, adopted \\
Dust yield (SNe Pop II) & $y_{\rm D,SN,PopII}$ & 4.0 $\times$ 10$^{-4}$ & \citep{ref33} \\
Grain-growth timescale at $0.1 Z_{\rm \odot}$ & $\tau_{\rm grow}(0.1 Z_\odot)$ & 4 Myr & This work; normalization chosen for efficient dense-ISM growth. $Z^{-1}$ scaling motivated by \citep{ref18}. \\
Grain-growth index & $n$ & 1 & This work; metallicity scaling motivated by \citep{ref18} \\
Grain-growth transition & $Z_{\rm gg,0} - Z_{\rm gg,1}$ & 0.08 - 0.12 Z$_{\rm \odot}$ & This work, explored during model development and fixed to $0.08 - 0.12 Z_{\rm \odot}$ in the final model. \\
Pop II transition metallicity & $Z_{\rm trans}$ & $10^{-4} Z_{\rm \odot}$ & \citep{ref34,ref35}; Population-II star-formation gate; used in the fiducial model and Pop III extensions. \\
Outflow loading factor & $\eta\ M_{\rm halo}$ & $\eta = {\rm clip} [(M_{\rm halo} / 10^{10} M_\odot)^{-0.5}, 0.02, 10]$ & \citep{ref36,ref37} \\
Initial halo mass & $M_{\rm halo,0}$ & $10^{7} - 10^{11}$ $M_{\rm \odot}$ & Explored; 11 halo masses \\
Halo growth timescale & $\tau_{\rm halo}$ & 100 Myr & \citep{ref36} \\
Pop III transition timescale & $\tau_{\rm PopIII,trans}$ & 10 Myr & Pop III extension only, fixed for robustness tests \\
Number of Pop III bursts & $N_{\rm burst}$ & 5 & Periodic Pop III extension only, fixed \\
Population III SFR amplitude & $SFR_{\rm III,peak}$ & 2 $\times$ $10^{-3}$ $M_{\rm \odot}$ yr${-1}$ & Pop III extension only, fixed for robustness tests \\
Continuous Pop III duration & $t_{\rm III,end}$ & 10 Myr & Continuous Pop III extension only, fixed \\
Population III history & $SFH_{\rm PopIII}$ & continuous early episode \& periodic 5-burst history & Robustness tests only \\
Pop III burst width & $\Delta_{\rm burst}$ & 0.2 Myr & Periodic Pop III extension only, fixed \\
Dust-destruction timescale & $\tau_{\rm dest}(SFR)$ & 20 - 350 Myr & \citep{ref18} \\
FUV opacity  (SN phase) & $\kappa_{\rm FUV,SN}$ & $10^3$ cm$^2$ g$^{-1}$ & \citep{ref12}  \\
FUV opacity (ISM phase) & $\kappa_{\rm FUV,ISM}$ & $10^5$ cm$^2$ g$^{-1}$ & \citep{ref13}  \\
Covering fraction & $f_{\rm cov}\ M_{\rm star}$ & 0.5000 - 0.9995 & \citep{ref12} \\
Radius normalization & $R_{\rm e,0}$ & 0.8 kpc & \citep{ref38,ref39} \\
Radius - mass slope & $\alpha$ & 0.20 & \citep{ref38} \\
Radius - redshift slope & $\beta$ & 0.30 & \citep{ref38} \\
Cosmic baryon fraction & $f_{\rm b}$ & 0.156 & \citep{ref40} \\
Inflow efficiency & $f_{in}$ & 1.0 & \citep{ref36,ref37} \\
Pre-transition metal-mixing normalization & $f_{\rm mix,pre}$ & 0.9607 & Calibrated once from the transition timing; fixed in the final model \\
Opacity source-balance midpoint & $f_{\rm source,50}$ & 0.9268 & Calibrated once from the attenuation response; fixed in the final model \\
Post-transition star-formation efficiency factor & $f_{\rm SF,post}$ & 1.00, 0.30, 0.10, 0.03 & Explored as four paired physical states \\
Low-metallicity return fraction & $R_{\rm lowZ}$ & 0.65, 0.75, 0.84, 0.90 & Explored as four paired physical states \\
Low-metallicity long-lived stellar lock-up fraction & $f_{\rm lock}$ & 0.35, 0.25, 0.16, 0.10 & Explored as four paired physical states \\
Effective surviving SNe-dust yield factor & $f_{\rm SN,surv}$ & 0.77, 1.30, 2.19, 3.68 $\times$ fiducial & Explored as four paired physical states \\
Dense-gas fraction & $f_{\rm dense}$ & 0.01, 0.03, 0.07, 0.15 & Explored as four paired physical states \\
Fraction of fresh metals injected into hot phase & $f_{\rm Z,hot}$ & 0.96 & This work, fixed \\
$S_{\rm low}$ hot-to-SF mixing timescale & $\tau_{\rm Z,mix,slow}$ & 40 - 1750 Myr & Halo-mass dependent, fixed functional form \\
Fast hot-to-SF mixing timescale & $\tau_{\rm Z,mix,fast}$ & 25 - 80 Myr & Halo-mass dependent, fixed \\
Fast-mixing activation & $Z_{\rm mix,0} - Z_{\rm mix,1}$ & 0.10 - 0.13 $Z_{\rm \odot}$ & This work, fixed \\
SN dust-release delay & $\tau_{\rm SN}$ & 10 Myr & This work, fixed \\
Maximum depletion target & $f_{\rm dep,max}$ & 0.82 & This work, fixed \\
SF-efficiency activation & $Z_{\rm SF,0} - Z_{\rm SF,1}$ & 0.11 - 0.15 $Z_{\rm \odot}$ & This work, fixed \\
Opacity metallicity gate & $Z_{\rm \kappa,0} - Z_{\rm \kappa,1}$ & 0.075 - 0.100 $Z_{\rm \odot}$ & This work, fixed \\
Opacity Hill index & q & 8 & This work, fixed \\
Halo-accretion decline time & $t_{\rm decline}$ & 600 Myr & This work, fixed \\
Halo-accretion decline width & $\Delta t_{\rm decline}$ & 75 Myr & This work, fixed \\
Late-time accretion floor & $f_{\rm late}$ & 0.05 & This work, fixed \\
Dust-destruction state factor & $f_{\rm dest,state}$ & 0.08 - 5 & This work, paired-state dependence \\
\end{longtable}


\begin{thebibliography}{51}

\bibitem[Angl{\'e}s-Alc{\'a}zar et al.(2017)]{Angles-Alcazar2017} 
Angl{\'e}s-Alc{\'a}zar, D., Faucher-Gigu{\`e}re, C.-A., Kere{\v{s}}, D., et al.\ 2017, MNRAS, 470, 4, 4698
\bibitem[Akaike(1974)]{ref25}
Akaike, H. 1974, IEEE Trans. Automat. Control, 19, 716
\bibitem[Algera et al.(2024)]{ref10}
Algera, H. S. B., et al. 2024, MNRAS, 533, 3098
\bibitem[Algera et al.(2025)]{ref16}
Algera, H. S. B., et al. 2025, arXiv e-prints, arXiv:2512.14486
\bibitem[Asano et al.(2013)]{ref18}
Asano, R. S., Takeuchi, T. T., Hirashita, H., \& Inoue, A. K. 2013,
Earth Planets Space, 65, 213
\bibitem[Bakx et al.(2026)]{ref15}
Bakx, T. J. L. C., et al. 2026, MNRAS, 546, staf2284
\bibitem[Bianchi et al.(2007)]{ref46}
Bianchi, S. \& Schneider, R. 2007, MNRAS 378, 973
\bibitem[Bocchio et al.(2016)]{ref47}
Bocchio, M., Marassi, S., Schneider, R., et al. 2016, A\&A, 587, A157
\bibitem[Bromm \& Loeb(2003)]{ref34}
Bromm, V., \& Loeb, A. 2003, Nature, 425, 812
\bibitem[Burgarella et al.(2025)]{ref11}
Burgarella, D., et al. 2025, A\&A, 699, A336
\bibitem[Burgarella et al.(2026)]{ref12}
Burgarella, D., et al. 2026, arXiv e-prints, arXiv:2605.09829
\bibitem[Calabrò et al.(2024)]{ref20}
Calabrò, A., et al. 2024, ApJ, 975, 245
\bibitem[Castellano et al.(2024)]{ref19}
Castellano, M., et al. 2024, ApJ, 972, 143
\bibitem[Chavez Ortiz et al.(2025)]{ref22}
Chavez Ortiz, O. A., et al. 2025, arXiv e-prints, arXiv:2511.03035
\bibitem[Cullen et al.(2024)]{ref17}
Cullen, F., et al. 2024, MNRAS, 531, 997
\bibitem[Dekel et al.(2013)]{ref36}
Dekel, A., Zolotov, A., Tweed, D., et al. 2013, MNRAS, 435, 999
\bibitem[Dib et al.(2015)]{ref42}
Dib, S., Piau, L., Mohanty, S., Braine, J., et al. 2011, MNRAS 415, 3439
\bibitem[Feldmann(2011)]{ref41}
Feldmann R., 2015, MNRAS, 449, 3274
\bibitem[Ferrara et al.(2025)]{ref13}
Ferrara, A., Pallottini, A., \& Sommovigo, L. 2025, A\&A, 694, A286
\bibitem[Gall \& Hjorth(2018)]{ref02}
Gall, C., \& Hjorth, J. 2018, ApJ, 868, 62
\bibitem[Grudic et al.(2018)]{ref43}
Grudić, M. Y., Hopkins, P. F., Faucher-Giguère, C.-A., et al. 2018, MNRAS, 475, 3511
\bibitem[Harikane et al.(2023)]{ref23}
Harikane, Y., Ouchi, M., Oguri, M., et al. 2023, ApJS, 265, 5
\bibitem[Hashimoto et al.(2019)]{ref06}
Hashimoto, T., et al. 2019, PASJ, 71, 71
\bibitem[Heger \& Woosley(2002)]{ref30}
Heger, A., \& Woosley, S. E. 2002, ApJ, 567, 532
\bibitem[Heger \& Woosley(2010)]{ref49}
Heger, A., \& Woosley, S. E. 2010, ApJ, 724, 341
\bibitem[Hirashita(2000)]{ref45}
Hirashita, H. 2000, PASJ 52, 585
\bibitem[Kawamata et al.(2018)]{ref39}
Kawamata, R., Ishigaki, M., Shimasaku, K., et al. 2018, ApJ, 855, 4
\bibitem[Laporte et al.(2017)]{ref05}
Laporte, N., Ellis, R. S., Boone, F., et al. 2017, ApJL, 837, L21
\bibitem[Lilly et al.(2013)]{ref37}
Lilly, S. J., Carollo, C. M., Pipino, A., Renzini, A., \& Peng, Y. 2013,
ApJ, 772, 119
\bibitem[Limongi \& Chieffi(2018)]{ref31}
Limongi, M., \& Chieffi, A. 2018, ApJS, 237, 13
\bibitem[Marassi et al.(2019)]{ref33}
Marassi, S., Schneider, R., Limongi, M., et al. 2019, MNRAS, 484, 2587
\bibitem[Narayanan et al.(2026)]{ref14}
Narayanan, D., et al. 2026, Open J. Astrophys., 9, 59986
\bibitem[Natta \& Panagia(1984)]{ref27}
Natta, A., \& Panagia, N. 1984, ApJ, 287, 228
\bibitem[Nomoto et al.(2006)]{ref50}
Nomoto, K., Tominaga, N., Umeda, H., Kobayashi, C., \& Maeda, K. 2006, NuPhA, 777, 424
\bibitem[Nozawa et al.(2003)]{ref51}
Nozawa, T., Kozasa, T., Umeda, H., Maeda, K., \& Nomoto, K. 2003, ApJ, 598, 785
\bibitem[Nozawa et al.(2007)]{ref32}
Nozawa, T., Kozasa, T., Habe, A., et al. 2007, ApJ, 666, 955
\bibitem[Nozawa et al.(2007)]{ref29}
Parente, M., Salvestrini, F., Granato, G. L., et al. 2026,
arXiv e-prints, arXiv:2603.04505
\bibitem[Planck Collaboration et al.(2020)]{ref40}
Planck Collaboration, et al. 2020, A\&A, 641, A6
\bibitem[Pozzi et al.(2021)]{ref07}
Pozzi, F., et al. 2021, A\&A, 653, A84
\bibitem[Robertson et al.(2024)]{ref24}
Robertson, B., et al. 2024, ApJ, 970, 31
\bibitem[Sarangi \& Cherchneff(2015)]{ref03}
Sarangi, A., \& Cherchneff, I. 2015, A\&A, 575, A95
\bibitem[Schneider(2006)]{ref35}
Schneider, R. 2006, New Astron. Rev., 50, 64
\bibitem[Schneider \& Maiolino(2024)]{ref01}
Schneider, R., \& Maiolino, R. 2024, A\&ARv, 32, 2
\bibitem[Schwarz(1978)]{ref26}
Schwarz, G. 1978, Ann. Stat., 6, 461
\bibitem[Shibuya et al.(2015)]{ref38}
Shibuya, T., Ouchi, M., \& Harikane, Y. 2015, ApJS, 219, 15
\bibitem[Sommovigo et al.(2022)]{ref28}
Sommovigo, L., et al. 2022, MNRAS, 513, 3122
\bibitem[Vincenzo et al.(2016)]{ref44}
Vincenzo, F., Matteucci, F., Belfiore, F. \& Maiolino, R. 2016, MNRAS 455, 4183
\bibitem[Wang et al.(2023)]{ref48} 
Wang, B., Fujimoto, S., Labb{\'e}, I., et al.\ 2023, ApJL, 957, L34
\bibitem[Watson et al.(2015)]{ref04}
Watson, D., Christensen, L., Knudsen, K. K., et al. 2015, Nature, 519, 327
\bibitem[Witstok et al.(2023)]{ref09}
Witstok, J., Jones, G. C., Maiolino, R., Smit, R., \& Schneider, R. 2023,
MNRAS, 523, 3119
\bibitem[Zavala et al.(2023)]{ref08}
Zavala, J. A., et al. 2023, ApJL, 943, L9
\bibitem[Zavala et al.(2024)]{ref21}
Zavala, J. A., et al. 2024, ApJL, 977, L9

\end{thebibliography}
\end{document}